# Transverse Quantum Confinement in Semiconductor Nanofilms: Optical Spectra and Multiple Exciton Generation

Vladimir I. Makarov[1] and Igor Khmelinskii[2]

[1]Department of Physics, University of Puerto Rico, Rio Piedras, P.O. Box 23343, San Juan, Puerto Rico 00931-3343, USA

[2]Universidade do Algarve, FCT, DQF, and CIQA, P8005-139, Faro, Portugal

*Corresponding author*: Dr. Vladimir Makarov

*Contact information*: Department of Physics
University of Puerto Rico, Rio Piedras Campus
PO Box 23343, San Juan PR 00931-3343, USA

*Phone*: 1(787)529-2010
*Fax*: 1(787)756-7717
*E-mail*: vmvimakarov@gmail.com

**Abstract**

We report absorption and photoluminescence spectra of Si and $SnO_2$ polycrystalline nanofilms in the UV-Vis-NIR range, featuring discrete bands resulting from transverse quantum confinement. The film thickness ranged 3.9 nm to 12.2 nm, depending on the material. The results are interpreted within the particle-in-a-box model, with the box width equal to the layer mass thickness. The energy levels and transitions scale as the inverse square of the film thickness. The calculated values of the effective electron mass are independent on the film thickness and equal to $0.17m_o$ (Si) and $0.21m_o$ ($SnO_2$), with $m_o$ the mass of the free electron. The uncertainties in the effective mass values are ca.

2.5%, determined by the film thickness calibration. The second calculated model parameter, the quantum number $n$ of the HOMO, was also thickness-independent: 8.00 (Si) and 7.00 (SnO$_2$). This indicates that the Fermi level should also scale as the inverse square of the film thickness in these nanofilms. The observed transitions all start at the level $n$ and correspond to $\Delta n$ = 1, 2, 3, etc. The photoluminescence bands exhibit large Stokes shifts, moving to higher energies with increased excitation energy. The photoluminescence quantum yields exceed unity, showing evidence of multiple exciton generation from each absorbed photon. A prototype Si-SnO$_2$ nanofilm photovoltaic cell demonstrated an increase of the photoelectron yield with the photon energy, showing evidence of multiple exciton generation.

**Introduction**

Lately, a lot of interest was created around quantum confinement (QC) effects in different materials, with numerous publications in this area; see for example [1]. Three-dimensional, two-dimensional and one-dimensional QC has been observed in quantum dots, quantum rods, and quantum films, respectively [1, 2]. Earlier we reported indirect evidence for the one-dimensional transverse quantum confinement (TQC) in Si nanofilms, inferred from the interpretation of the exchange anticrossing spectra in nanosandwich structures [3]. Presently we report direct evidence for TQC in Si and $SnO_2$ nanofilms, in the form of UV-Vis-NIR absorption and emission spectra.

**Experimental**

Fused silica substrates 25 mm in diameter and 1 mm thick (Esco Optics) were used to deposit the films. Commercial Si and $SnO_2$ (Sigma/Aldrich) were used to produce nanocrystalline films on a commercial sputtering/thermo-evaporation Benchtop Turbo deposition system (Denton Vacuum). The substrate temperature was 475ºC. The film thickness was controlled by XRD [4], with the XPert MRD system (PANalytic) calibrated by standard nanofilms of the same materials. The estimated absolute uncertainty of film thickness was 2.5%; the relative uncertainties were much smaller, determined by the shutter opening times of the deposition system.

Absorption and emission spectra were recorded on Hitachi U-3900H UV-Visible Spectrophotometer and Edinburgh Instruments FS5 Spectrofluorometer. The absorption spectra in the near-IR were recorded on a PF 2000 FTIR spectrometer (Perkin Elmer). The spectral peak maxima were located using the PeakFit software (Sigmaplot).

Polynomials were fitted and fitting uncertainties estimated using the LINEST function in Excel (Microsoft).

Photoelectric current response measurements were carried out using a high-pressure Xe lamp ($W$ = 1000 W; Ariel Corporation, Model 66023), a monochromator (Thermo Jarrell Ash, Mono Spec/50), a model 2182A nanovoltmeter (Keithley Instruments) connected to a computer by GPIB interface, and home-made software in the LabView programming environment (National Instruments).

Low-temperature measurements were carried out at 77 K using an Optistat DN-V2 optical cryostat (OXFORD Instruments). All measurements were made at 77 K, unless expressly stated otherwise.

**Results and discussion**

**Si nanofilms**

Si nanofilms at 5.7, 8.1, 10.3 and 12.2 nm were deposited on fused silica substrates. Figure 1 shows UV-Vis spectra for 5.7 and 10.3 nm films, and the NIR spectrum of the 5.7 nm film. The band maxima are listed in Table 1.

<Insert Figure 1>

Table 1. UV-Vis-NIR absorption band maxima (cm$^{-1}$) and the $E_1$ fitting parameter (with fitting uncertainty) in Si nanofilms of different thickness. Predicted bands are listed in italics. The $\Delta n$ values were attributed as explained in the text.

|  | Film thickness, nm | | | |
|---|---|---|---|---|
|  | 5.7 | 8.1 | 10.3 | 12.2 |
| $E_1$, cm$^{-1}$ | 542.0±0.2 | 268.50±0.06 | 166.04±0.02 | 118.31±0.02 |
| $\Delta n$ | Absorption band maxima, cm$^{-1}$ | | | |
| 1 | 9216 | *4564* | *2823* | *2010* |
| 2 | 19516 | *9664* | *5977* | *4259* |
| 3 | 30901 | 15302 | *9463* | *6744* |
| 4 | 43369 | 21476 | 13282 | *9466* |
| 5 | | 28188 | 17432 | 12425 |
| 6 | | 35436 | 21915 | 15620 |
| 7 | | 43222 | 26730 | 19052 |
| 8 | | 51544 | 31876 | 22721 |
| 9 | | | 37355 | 26626 |
| 10 | | | 43166 | 30768 |
| 11 | | | 49309 | 35146 |
| 12 | | | | 39762 |
| 13 | | | | 44613 |
| — | | | | 51713 |

The transition energies were interpreted using the particle-in-a-box model with infinite walls. The energy levels in such a system are given by the following equation, quadratic in the quantum number $n$:

$$E_n = \frac{h^2 n^2}{8 m^* L^2} \quad (1)$$

here $h$ is the Planck constant, $m^*$ the effective electron mass and $L$ the box width (nanofilm thickness). Thus for the transition energies we obtain, introducing the residual $\delta E$ to compensate for the experimental errors:

$$E_{n+\Delta n} - E_n = E_1\left(\Delta n^2 + 2n\Delta n\right) + \delta E \quad (2)$$

with

$$E_1 = \frac{h^2}{8 m^* L^2} \quad (3)$$

We fitted the second-order polynomial of Eq. (2) to the series of measured band maxima, for each of the samples separately. We obtained $1-R^2 < 6\times 10^{-10}$ and statistical uncertainties for the $E_1$ and $2nE_1$ coefficients below 0.04%. Note that the last band in the 12.2 nm sample was an obvious outlier by about 2000 cm$^{-1}$, probably due to interaction with an atomic level, and therefore was not used in the fit. The $\Delta n$ value for the lowest-energy recorded band in each sample was chosen so as to minimize the respective $\delta E$ absolute values, which were statistically insignificant. Using the respective $E_1$ values and the film thicknesses, we calculated the effective electron mass $m^* = 0.17m_o$, with differences below 0.1% between samples, demonstrating the very low relative uncertainties in the film thickness. These values are similar to those obtained for electrons and light holes in bulk Si. The same fits produced the starting quantum number $n = 8.00$ in all samples. This quantum number corresponds to the HOMO orbital of the discrete transverse-quantized level system, which should lie below the Fermi level to be filled. Apparently the $\Delta n = \pm 1$ selection rule of the particle-in-a-box model does not apply any more; we interpret this as an indication that the transitions are occurring between adjacent atomic chains that have sufficient overlap of their wavefunctions, which also explains why the transitions polarized in the film plane are allowed, while only the transverse-polarized transitions would be allowed in the original particle-in-a-box model. We explain the absence of the transitions from the $n$-1 *etc*. states by the lack of overlap of the respective stronger-binding wavefunctions with those in the adjacent atomic chains. In any case, we expect that future rigorous quantum mechanical calculations will produce adequate *ab initio* values of the transition energies and transition strengths in nanofilms.

Absorption spectra of similar Si films were investigated previously, using ellipsometry [5]. However, the published absorption spectra of polycrystalline Si films in the 5 to 30 nm thickness range have little in common with the presently obtained data, probably due to undetected difficulties in solving the inverse ellipsometric problem as part of the spectral reconstruction [6].

Apparently, the simple model we use is describing the energy levels and transitions with remarkable precision, as we shall also see in $SnO_2$ films.

We recorded photoluminescence (PL) spectra of the Si films at 77K, see Fig. 2. The PL bands shift to higher energies when higher-energy excitation was used. Note the concurrent increase in relative intensity, indicative of multiple exciton generation [8]. Note also the presence of another, probably more intense band, as indicated by the tail at ca. 12000 cm$^{-1}$. The maximum of this band should be below 9216 cm$^{-1}$, the maximum of the first absorption band in this film. However, we were unable to record PL spectra in the NIR.

<Insert Figure 2>

**Si data and the between-zone transition model of quantum-well spectral properties**

The between-zone transition model successfully used for multilayer quantum-well semiconductor structures predicts quadratic dependence of the transition onset energy on the quantum number (with $\Delta n = 0$ selection rule) and the spectral intensities that generally grow stepwise, as each of the successive transitions steps in with growing photon energy [7]. In this model, the effect of the film thickness manifests itself in the shifts of the hole and electron levels in the respective valence and conduction zones of the semiconductor, and in adding more steps into the spectral curve within the fixed energy interval. These spectral shifts are relatively weak due to the large total thickness of the multilayer quantum well structure (hundreds of nanometers), translating into equally weak shifts of the absorption onsets and peaks attributable to holes and electrons. In such thick systems the behavior is essentially bulk-like, with the transverse quantization effects being relatively small and producing only minor band shifts. These between-zone transitions are clearly apparent in the spectra of Si films (Figure 1), as the background with the intensity rise towards higher energies, similar to the spectrum of bulk Si. In contrast, the bands we observe in single-layer nanometer-thin films have a linear term in their dependence on the quantum number (in fact, on the quantum number change) complementing the quadratic term, see Eq. (2). These bands suffer major shifts, moving from the UV into the IR with the growing film thickness, going well below the absorption onset of bulk Si. The growing band width and the decreasing band spacing make the neighboring absorption bands indistinguishable in films of about 20-30 nm thick (compare the spectra of the two films in Figure 1), and the spectrum becomes similar to that predicted by the between-zone transition model, as the bulk-like zone structure is formed along the transverse coordinate and the model becomes capable of interpreting all of the observed spectral features. Apparently the bulk-Si peaks attributable to electrons and holes are not resolved or too weak in our experimental conditions. The peak series that we observe, however, shows that the properties of very thin nanofilms are dominated by transverse quantum confinement effects, as opposed to thicker films, where such effects are only a weak perturbation.

**$SnO_2$ nanofilms**

$SnO_2$ nanofilms 3.9, 5.8 and 8.9 nm thick were deposited on fused silica substrates. Figure 3 shows selected UV-Vis absorption spectra, and Table 2 lists the absorption band maxima.

<Insert Figure 3>

Table 2. UV-Vis absorption band maxima (cm$^{-1}$) and the $E_1$ fitting parameter (with fitting uncertainty) in SnO$_2$ nanofilms of different thickness. Predicted bands are listed in italics. The $\Delta n$ values were attributed as explained in the text.

|  | Film thickness, nm | | |
|---|---|---|---|
|  | 3.9 | 5.8 | 8.9 |
| $E_1$, cm$^{-1}$ | 970.5* | 438.8±0.3 | 186.38±0.03 |
| $\Delta n$ | Absorption band maxima, cm$^{-1}$ | | |
| 1 | 14558 | *6582* | *2795* |
| 2 | 31057 | 14042 | *5963* |
| 3 | 49497 | 22379 | *9504* |
| 4 |  | 31595 | 13418 |
| 5 |  | 41687 | 17704 |
| 6 |  | *52657* | 22363 |
| 7 |  |  | 27395 |
| 8 |  |  | 32800 |
| 9 |  |  | 38577 |
| 10 |  |  | 44727 |

*No uncertainty estimate available, as only 3 bands were fitted (exactly) by the second-order polynomial.

Similar to Si samples, these band series were fitted using Eq. (2), and the starting $\Delta n$ values were chosen so as to minimize the absolute values of $\delta E$ for each of the samples, which were statistically insignificant (apart from the 3.9 nm sample, where no error estimate was available). Once again, very good quality fits were obtained, producing the HOMO quantum number $n = 7.00$ and the effective electron mass $m^* = 0.21 m_o$, with very small relative film thickness uncertainty. Note that $m^*$ values quoted for bulk SnO$_2$, $m^* = (0.23 \ldots 0.30) m_o$, are similar to the one obtained here [9].

**Nanofilm photovoltaic cell**

We investigated a prototype photovoltaic cell with multiple exciton generation based on Si and SnO$_2$, schematically shown in Fig. 4. Figure 5 shows the UV-Vis absorption spectra of separate layers, and the spectrum of the entire cell, before the thick Au layer was deposited.

< Insert Figure 4>

<Insert Figure 5>

Note that the UV-Vis absorption spectra of Si and the $SnO_2$ layers in the cell show the same absorption bands as the respective isolated nanofilms, demonstrating that transverse quantization in semiconductor nanofilms operates independently in each of the stacked layers, at least for the two layers of different composition. We recorded the excitation wavelength dependence of the photocurrent at 77K, with the results shown in Figure 6, normalized to the number of exciting photons.

<Insert Figure 6>

We see that excitation into the Si transverse-quantized bands produces a much stronger photocurrent than the excitation into the $SnO_2$ bands, while the excitation outside of these bands produces very low photocurrent. Note that the quantum yield of photoelectrons is roughly proportional to the photon energy. This fact and the quantum yields of photoelectrons exceeding unity ($\phi$ = 2.5 electrons/photon when exited into the short-wavelength Si band at 37244 cm$^{-1}$) are clear indicators of the multiple exciton generation in this photovoltaic cell [8]. The open-circuit voltage of the cell was 35.7 mV, which seems reasonable for such a small film thickness.

**Conclusion**

We report transverse one-dimensional quantum confinement in semiconductor nanofilms, showing that a simple particle-in-a-box model adequately describes the structure of the electronic levels quantized in the transverse direction. We show that optical spectroscopy may be used to precisely measure the nanofilm thickness in the range dependent on the material, provided the respective effective electron mass is known. We also demonstrated multiple exciton generation in a prototype Si-SnO$_2$ photovoltaic cell at 77 K. These findings provide a new understanding of the physics of nanofilms, opening a new range of possibilities for the technology of solid-state devices.

**Conflicts of interest**

The authors report no conflicts of interest.

**Figure captions**

Figure 1. UV-Vis spectra for 5.7 and 10.3 nm Si films, and the NIR spectrum of the 5.7 nm film.

Figure 2. Photoluminescence spectra of the 5.7 nm Si film at 77K. 1 - exctited into the 19516 cm$^{-1}$ band; 2 - excited into the 30901 cm$^{-1}$ band; 3 - excited into the 43369 cm$^{-1}$ band.

Figure 3. UV-Vis spectra for 3.9 and 5.8 nm $SnO_2$ films. The 3.9 nm film spectrum is shifted upwards by 0.02 O.D. units, for visual separation.

Figure 4. The prototype photovoltaic cell: 1 – fusel silica substrate (1 mm thick, 25 mm diameter); 2 – Au film (21.3 nm); 3 – Si film (3.9 nm); 4 – $SnO_2$ film (4.0 nm); 5 – Au film (0.109 μm); 6 – Cu plate (0.75 mm); 7 – hollow plastic cylinder; 8 – Cu ring. Note that the spectrum of the cell is the sum of spectra of its separate layers deposited on a fused silica substrate.

Figure 5. UV-Vis spectra: 1- the entire prototype photovoltaic cell, before the thick back Au layer was deposited; 2- Si layer; 3- the front Au layer; 4- $SnO_2$ layer. The spectra were shifted vertically, for visual separation.

Figure 6. Photoelectric current in the prototype photovoltaic cell normalized to the number of absorbed photons in function of the photon energy. The larger current peaks correspond to the excitation into the bands of the Si nanofilm, and the smaller – into those of the SnO$_2$ nanofilm.

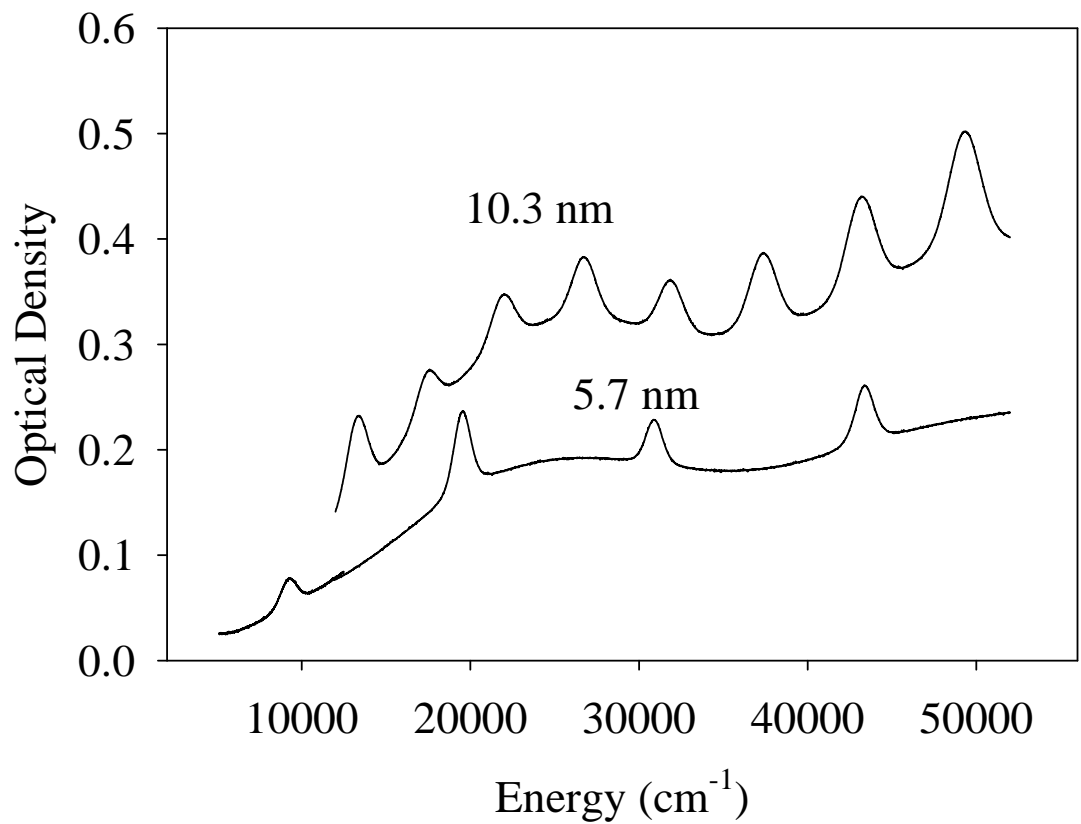

Figure 1; Makarov, Khmelinskii

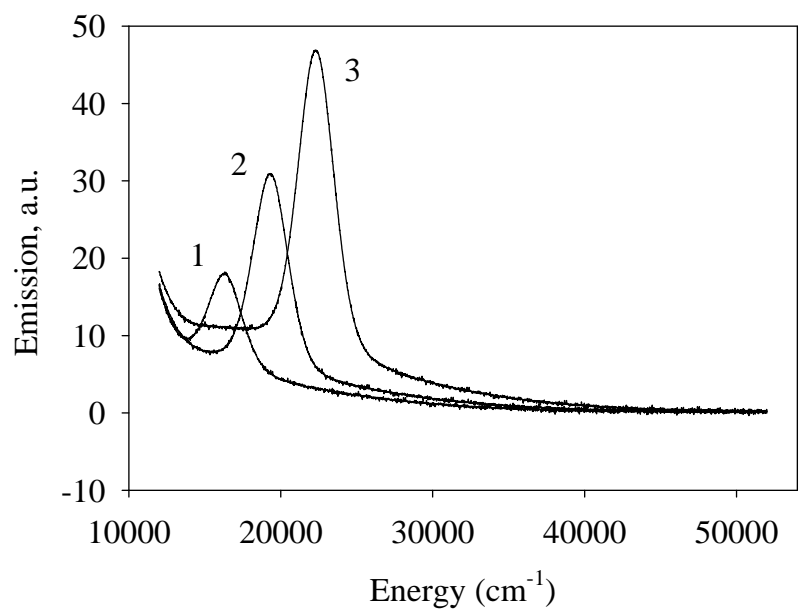

Figure 2; Makarov, Khmelinskii

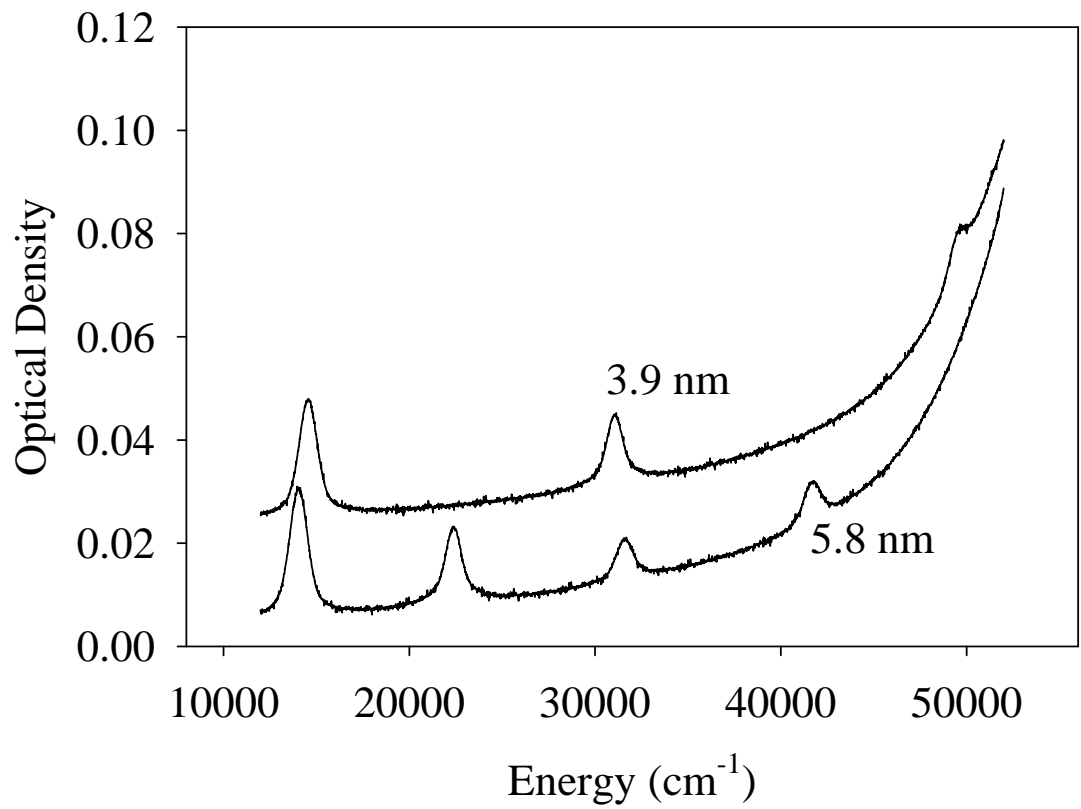

Figure 3; Makarov, Khmelinskii

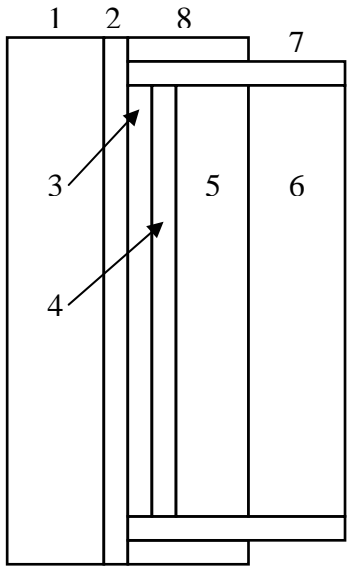

Figure 4; Makarov, Khmelinskii

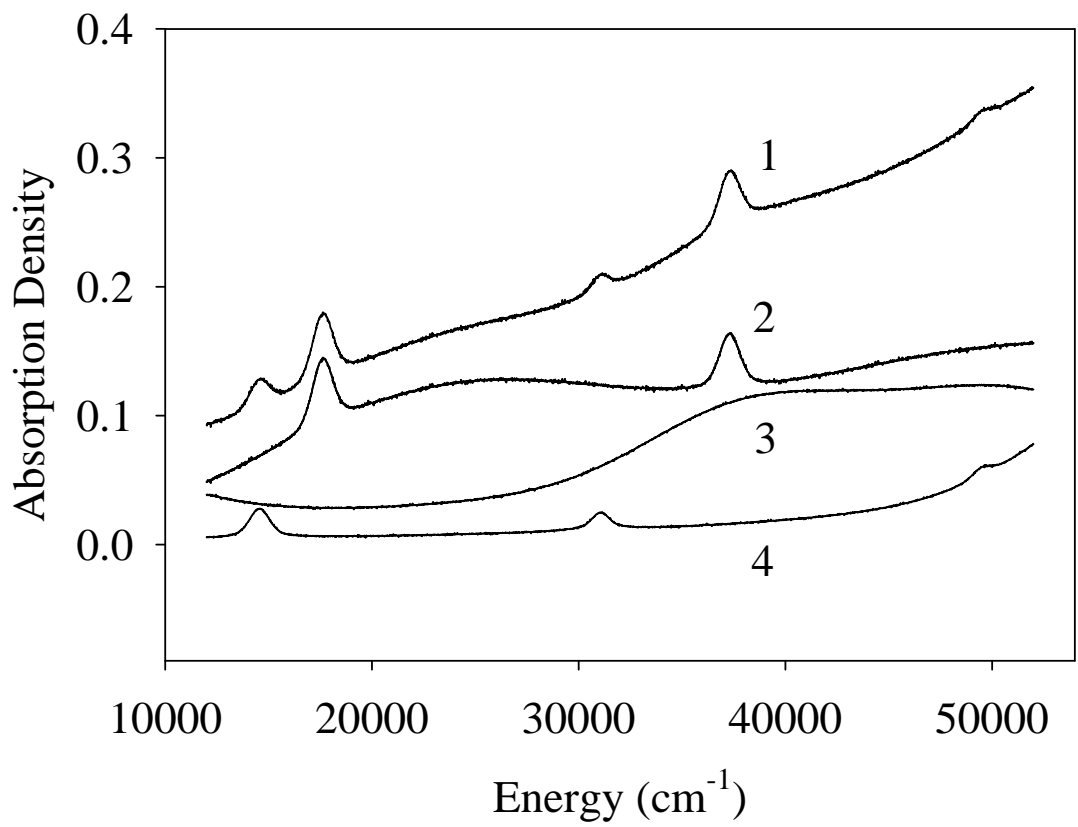

Figure 5; Makarov, Khmelinskii

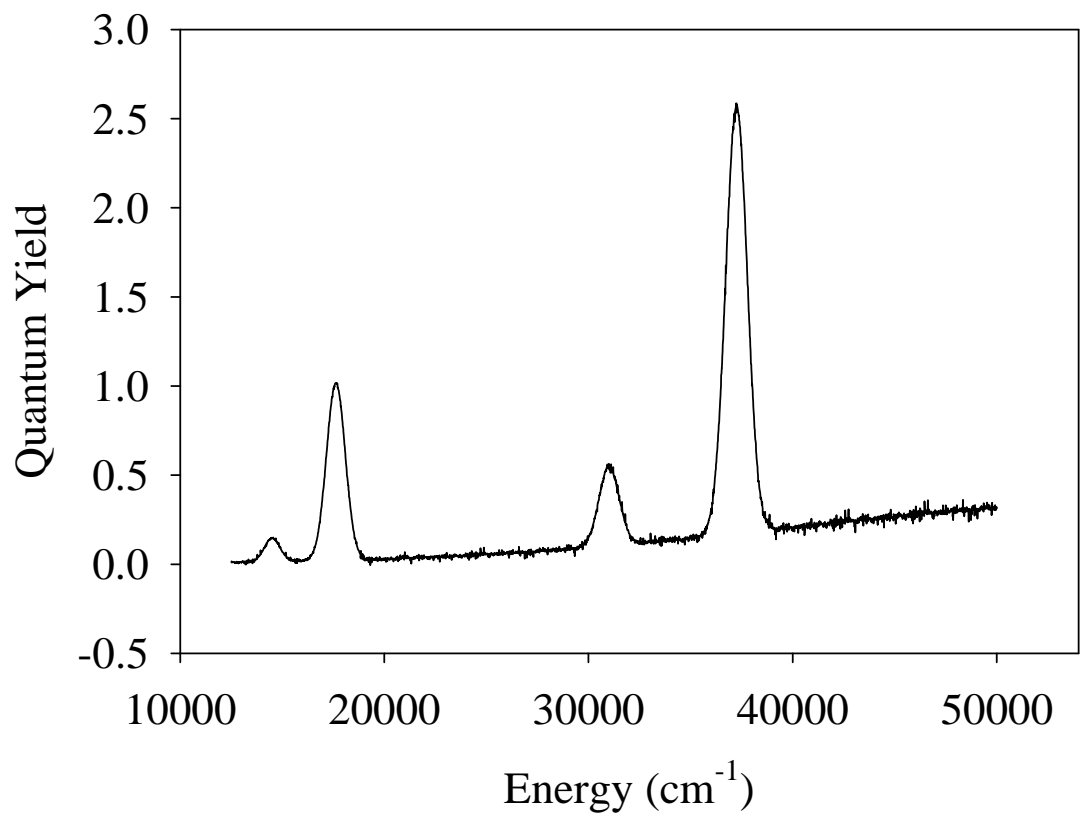

Figure 6; Makarov, Khmelinskii